\documentclass[prl,aps,twocolumn,showpacs,superscriptaddress,notitlepage,footinbib]{revtex4-1}

\usepackage{graphicx}
\usepackage{amsmath}
\usepackage{amssymb}
\usepackage{times}
\usepackage{color}
\usepackage{bbm}
\usepackage{bm}
\usepackage{xcolor}
\usepackage[colorlinks, citecolor=blue, linkcolor=violet, breaklinks]{hyperref}

\usepackage[colorinlistoftodos, shadow,textsize= footnotesize]{todonotes}

\newcommand{\vect}[1]{\bm{#1}}

\newcommand{\be}{\begin{equation}}
\newcommand{\ee}{\end{equation}}
\newcommand{\ud}{\mathrm{d}}
\newcommand{\beq}{\begin{eqnarray}}
\newcommand{\eeq}{\end{eqnarray}}

\begin{document}

\title{Spin-Mixing Interferometry with Bose-Einstein Condensates}

\author{Marco Gabbrielli}
\affiliation{Dipartimento di Fisica e Astronomia, Universit\`a degli Studi di Firenze, via Sansone 1, I-50019, Sesto Fiorentino, Italy}
\affiliation{QSTAR, INO-CNR and LENS, Largo Enrico Fermi 2, I-50125 Firenze, Italy}

\author{Luca Pezz\`e}
\email{luca.pezze@ino.it}
\affiliation{QSTAR, INO-CNR and LENS, Largo Enrico Fermi 2, I-50125 Firenze, Italy}

\author{Augusto Smerzi}
\affiliation{QSTAR, INO-CNR and LENS, Largo Enrico Fermi 2, I-50125 Firenze, Italy}

\begin{abstract}
Unstable spinor Bose-Einstein condensates are ideal candidates to create nonlinear three-mode interferometers. 
Our analysis goes beyond the standard SU(1,1) parametric approach and therefore provides the regime of parameters where 
sub-shot-noise sensitivities can be reached with respect to the input total average number of particles.
Decoherence due to particle losses and finite detection efficiency are also considered. 
\end{abstract}

\date{\today}


\maketitle

Interferometers provide the most precise measurements in physics~\cite{note0, HariharanBOOK,CroninRMP2009,*SchnabelNATCOMM2010}.
Hence, there is an urgent demand for novel theoretical proposals and experimental 
techniques aimed at further increasing their sensitivity. 
Most of the current atomic and optical interferometers are made of linear devices such as beam splitters and phase shifters. 
Their phase uncertainty is fundamentally bounded by the shot-noise limit $\Delta \theta  \sim 1/\sqrt{\bar{n}}$, 
when using probe states made of average $\bar{n}$ uncorrelated particles~\cite{PezzePRL2009, GiovannettiPRL2006}.
It has been clarified that overcoming this bound requires engineering proper particle-entangled states~\cite{PezzePRL2009}
(see~ Refs. \cite{GiovannettiNATPHOT2011, Toth, Varenna} for reviews). 
Using such states, sub-shot-noise (SSN) phase uncertainties have been demonstrated in several recent 
proof-of-principle experiments with atoms~\cite{GrossNATURE2010,OckeloenPRL2013,MusselPRL2014,StrobelSCIENCE2014,LuckeSCIENCE2011, 
LerouxPRL2010,*AppelPNAS2009,*BohnetNATPHOT2014} 
and photons~\cite{NagataSCIENCE2007,*XiangNATPHOT2011,*KrischekPRL2011,*KacprowiczNATPHOT2010,*AfekSCIENCE2010}. 
Yet, noise and decoherence limit the creation and use of quantum correlations~\cite{EscherNATPHYS2010,*DemkowiczNATCOMM2012}.
It is therefore crucial to search for alternative schemes 
where probe states are classical and quantum correlations useful to reach SSN sensitivities are 
created {\it inside} the interferometer~\cite{GrossNATURE2010,RiedelNATURE2010,OckeloenPRL2013,MusselPRL2014}. 

\begin{figure}[b!]
\includegraphics[width=0.46\textwidth]{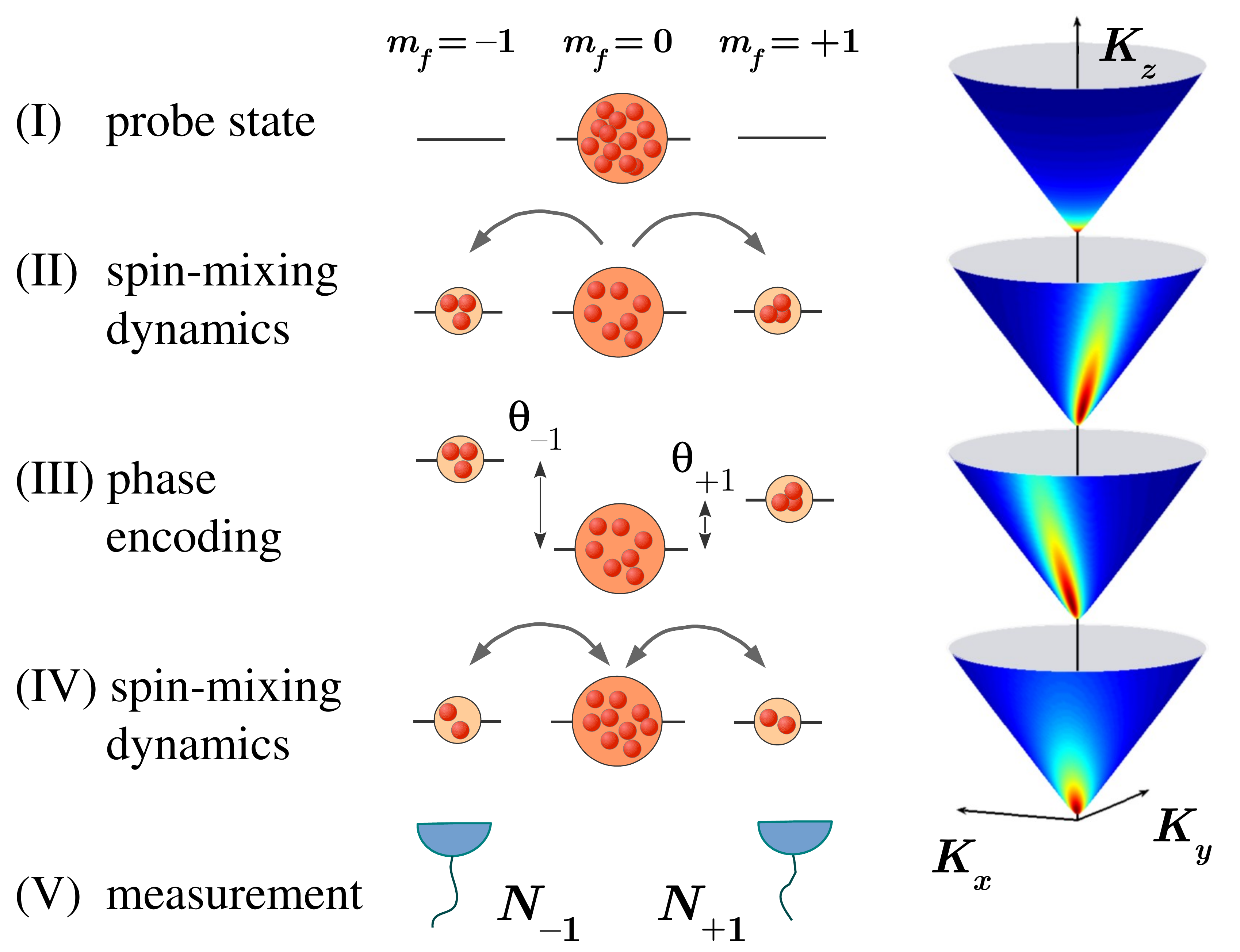}
\caption{%(color online) 
Left: scheme of spin-mixing interferometry with spinor BEC, here represented in the $f=1$ manifold. 
Right: when the $m_f=0$ mode is treated parametrically (mean-field approach), 
the interferometer operations can be 
visualized  %in the $(K_x, K_y, K_z)$ space~\cite{YurkePRA1986} 
on a hyperbolic surface by projecting the transformed state over SU(1,1) coherent states~\cite{YurkePRA1986, BrifJPA1996}.
} 
\label{fig1}
\end{figure}

In this Letter, we show that the coherent spin-mixing dynamics~(SMD) in 
a spinor Bose-Einstein condensate (BEC)~\cite{HoPRL1998,*OhmiPRL1998, Stamper-KurnRMP2013} 
can be exploited to realize a nonlinear three-mode interferometer, as shown in Fig.~\ref{fig1}. 
The SMD consists of binary collisions 
that coherently transfer correlated pairs of trapped atoms with opposite magnetic moment~\cite{PuPRL2000,*DuanPRL2000} 
from the $m_f=0$ to the $m_f=\pm1$ hyperfine modes, and vice versa. 
The probe state of the interferometer is classical, given by a condensate initially prepared in the $m_f=0$ mode, 
and quantum correlations are created by the~SMD.
We first study the interferometer in the mean-field limit, 
the $m_f=0$ mode operator being replaced by a $c$-number.
This analysis is valid for a large number of particles and low transfer rates.
In this case, the interferometer operations belong to the SU(1,1) group and it is possible to obtain
analytical predictions for the phase sensitivity.
In optical systems, where transfer rates are rather low, the probe state needs to be very intense and 
the SU(1,1) approach is well justified~\cite{YurkePRA1986}. 
SU(1,1) optical interferometry has been theoretically discussed~\cite{YurkePRA1986, PlickNJP2010, OuPRA2012, MarinoPRA2012}
and recently experimentally realized~\cite{HudelistNATCOMM2014}.
In contrast, experiments with spinor BECs~\cite{LuckeSCIENCE2011, GrossNATURE2011, HamleyNATPHYS2012,*GervingNATCOMM2013} 
can be performed well outside the mean-field regime,
with probe states of a relatively small number of particles and -- thanks to strong nonlinearities -- comparatively high transfer rates.
We have thus also implemented a full three-mode quantum analysis. 
Within this framework, we can rigorously provide phase sensitivity bounds 
with respect to the average total number of particles $\bar{n}$ in input. 
For realistic values of $\bar{n}$, including particle losses and finite detection efficiency, SSN is obtained in a regime 
where quantum corrections to the mean-field picture are important.  

%%%%%%%%%%%%%%%%%%%%%%%%%%%%%%%%%%%%%%%%%%%%%%%%%%%%%%%%%%%%%%%
% figure 1
%%%%%%%%%%%%%%%%%%%%%%%%%%%%%%%%%%%%%%%%%%%%%%%%%%%%%%%%%%%%%%%
%%%%%%%%%%%%%%%%%%%%%%%%%%%%%%%%%%%%%%%%%%%%%%%%%%%%%%%%%%%%%%%
%%%%%%%%%%%%%%%%%%%%%%%%%%%%%%%%%%%%%%%%%%%%%%%%%%%%%%%%%%%%%%%
%%%%%%%%%%%%%%%%%%%%%%%%%%%%%%%%%%%%%%%%%%%%%%%%%%%%%%%%%%%%%%%

{\it Spin-mixing interferometry with BECs.}~The protocol outlined in Fig.~\ref{fig1} follows five steps:
({I}) probe state preparation -- 
we consider empty $m_f=\pm 1$ modes and a BEC of average $\bar{n}$ atoms in the $m_f=0$ mode,
%described by a coherent state of average population $\bar{n}$ --
({II}) a first~SMD, ({III}) phase encoding,
and ({IV}) a second SMD.
Finally ({V}) the atoms are released from the trap: the three magnetic modes are spatially 
separated and the particle number is measured by imaging the atomic clouds.

A standard description of the~SMD is obtained in the single-mode approximation~\cite{KawaguchiPHYSREP2012}:
the condensate spatial wave function $\psi(\vect{r})$ in the $m_f= \pm 1$ modes is assumed to be the same as in the $m_f=0$ mode
and it is given by the solution of the Gross-Pitaevskii equation in the trapping potential~\cite{LawPRL1998}.
This approximation is justified for a relatively low atom number $\bar{n} \lesssim 10^5$, and tight confinement, 
when the spin healing length is larger than the size of the atomic cloud. These 
conditions are fulfilled in typical experimental setups~\cite{Stamper-KurnRMP2013}.
The field operators are thus approximated by $\hat{\Psi}_{i}(\vect{r}) = \psi(\vect{r}) \hat{a}_{i}$, 
where $\hat{a}_{i}$ ($\hat{a}^\dag_{i}$) are annihilation (creation) operators for modes $i=m_f=0, \pm 1$ obeying the 
boson commutation relations $[\hat{a}_i, \hat{a}_j^{\dag}]=\delta_{i,j}$ ($\hat{N}_i = \hat{a}_i^\dag \hat{a}_i$ 
is the particle number operator).
Up to terms proportional to the constant total particle number $\hat{N}=\hat{N}_{-1}+\hat{N}_{0}+\hat{N}_{+1}$, 
the many-body Hamiltonian describing the SMD in a dilute atomic cloud is~\cite{LawPRL1998} 
\begin{eqnarray} \label{Hamiltonian}
\hat{H}_{\rm SMD} &=&
 \chi \hbar \, ( e^{2i \phi} \hat{a}_0^{\dag} \hat{a}_0^{\dag} \hat{a}_{+1} \hat{a}_{-1} + e^{-2i \phi}
  \hat{a}_0 \hat{a}_0 \hat{a}_{+1}^{\dag} \hat{a}_{-1}^{\dag} ) + \nonumber \\
  & + & \chi \hbar \, \big(\hat{N}_0-\tfrac{1}{2} \big) ( \hat{N}_{+1} +  \hat{N}_{-1}).
\end{eqnarray}
The first term is identical to four-wave mixing in nonlinear optics~\cite{YurkePRA1986, ScullyBOOK}, 
where $\phi$ is the relative phase between the $m_f=0$ and $m_f=\pm 1$ modes. 
The second term in Eq.~(\ref{Hamiltonian}) is a mean-field shift. 
The coupling $\chi= \frac{4\pi \hbar}{3 M} (c_2 - c_0) \int d^3 \vect{r} \vert \psi(\vect{r}) \vert^4$
depends on the $s$-wave scattering lengths $c_0$ and $c_2$  of two bosons of mass $M$
scattering in the total spin channels $F=0$ and $F=2$, respectively~\cite{HoPRL1998,OhmiPRL1998,WideraNJP2006}.
We indicate as $(\chi t, \phi)_1$ [$(\chi t, \phi)_2$] the parameters for the 
first [second]~SMD. 
Experimentally, the~SMD can be accurately controlled via microwave dressing~\cite{Stamper-KurnRMP2013} and, in particular, switched off
during phase acquisition. 
Neglecting interaction between particles
during this stage, the (linear) phase shift Hamiltonian is 
\be \label{HPS}
\hat{H}_{\rm PS} =  \hbar q \big(\hat{N}_{+1}+\hat{N}_{-1}\big),
\ee
where $\hbar q$ is the energy difference between the $m_f=0$ and the $m_f=\pm 1$ modes, see Fig.~\ref{fig1}.
The unitary transformation $e^{-i \hat{H}_{\rm PS} t_{\rm PS} / \hbar}$
encodes the phase shift $\theta=\theta_{+1}+\theta_{-1} = 2qt_{\rm PS}$, 
where $\theta_{\pm1}$ are the phases accumulated by the atoms in the $m_f=\pm1$ modes, relative to the ones in the $m_f=0$ mode,
during a time $t_{\rm PS}$.
For instance, the signal can be the second-order Zeeman shift due to a sufficiently strong magnetic field.
Note that the first-order Zeeman shift, proportional to the net magnetization 
(equal to zero for our initial state), is conserved.

The phase shift is estimated by measuring the number of particles in the $m_f=\pm1$ modes at the end of the interferometric sequence.
We calculate the phase uncertainty as $\Delta \theta_{\rm CR} = 1/\sqrt{m F(\theta)}$, 
the Cram\'er-Rao lower bound~\cite{HelstromBOOK1976,GiovannettiNATPHOT2011, Varenna}, where $m$ accounts for 
the repetition of independent measurements, 
\be \label{Fisher}
F(\theta) \equiv \sum_{N_{\pm 1}=0}^{\infty} \frac{1}{P(N_{\pm 1}\vert \theta)} \bigg( \frac{\ud P(N_{\pm 1}\vert \theta)}{\ud \theta}\bigg)^2
\ee
is the Fisher information (FI) and $P(N_{\pm 1}\vert \theta)$ is the conditional probability to 
measure $N_{\pm1}$ particles given the phase shift $\theta$.
$\Delta \theta_{\rm CR}$ is a saturable lower bound of phase uncertainty~\cite{GiovannettiNATPHOT2011,HelstromBOOK1976, Varenna}. 
The FI can be experimentally extracted following the method demonstrated in~\cite{StrobelSCIENCE2014}.
Alternatively, we can calculate the phase uncertainty from the error 
propagation, $\Delta \theta_{\rm ep} = \frac{(\Delta \hat{N}_{\pm 1})_{\rm out}}{ \vert\ud \langle \hat{N}_{\pm 1} \rangle_{\rm out}/\ud\theta \vert}$, where 
$\langle \hat{N}_{\pm 1} \rangle_{\rm out}$ is the average number of particles in output
%the output \new{$m_f=\pm1$ modes} of the interferometer 
and $(\Delta \hat{N}_{\pm 1})^2_{\rm out}$ is the corresponding variance.
This method is experimentally feasible but not always optimal: we have $\Delta \theta_{\rm ep}/\sqrt{m} \geq \Delta \theta_{\rm CR}$, in general~\cite{PezzePRL2009,Varenna}.

%%%%%%%%%%%%%%%%%%%%%%%%%%%%%%%%%%%%%%%%%%%%%%%%%%%%%%%%%%%%%%%

{\it Mean-field approach.}~When the initial condensate contains a large number of particles and is weakly affected by the~SMD, 
we can study the interferometer operations by replacing $\hat{a}_0$ with $\sqrt{\bar{n}}$. 
We introduce the operators 
$\hat{K}_x = \frac{1}{2}(\hat{a}_{+1}^\dag \hat{a}_{-1}^\dag + \hat{a}_{+1}\hat{a}_{-1})$, 
$\hat{K}_y = \frac{1}{2i}(\hat{a}^\dag_{+1} \hat{a}_{-1}^\dag - \hat{a}_{+1}\hat{a}_{-1})$,
$\hat{K}_z = \frac{1}{2}(\hat{a}^\dag_{+1} \hat{a}_{+1} + \hat{a}_{-1}^{\dag}\hat{a}_{-1}+1)$,
which belong to the SU(1,1) group and satisfy
$[\hat{K}_x, \hat{K}_y]=-i \hat{K}_z$,
$[\hat{K}_y, \hat{K}_z]=i \hat{K}_x$ and $[\hat{K}_z, \hat{K}_x]=i \hat{K}_y$~\cite{YurkePRA1986,WodkiewiczJOSAB1985}. 
Equations (\ref{Hamiltonian}) and (\ref{HPS}) thus become, up to a constant term, 
$\hat{H}_{\rm SMD} = (2\bar{n}-1)\chi \hbar \hat K_z + 2\bar{n}\chi \hbar (\hat K_x \cos 2\phi + \hat{K}_y \sin 2\phi)$ and $\hat{H}_{\rm PS} = 2\hbar q \hat{K}_z$,
respectively.
%The \new{interferometer operations} can be visualized on a hyperbolic surface by projecting 
%the transformed state over SU(1,1) coherent states~\cite{BrifJPA1996}
%%% footnote
%%%\footnote{The Husimi distribution is obtained by projecting 
%%%over SU(1,1) coherent states \unexpanded{$| \beta, \phi \rangle = e^{\xi \hat{K}_+ - \xi^* \hat{K}_-} | 0 \rangle$}, 
%%%where \unexpanded{$\xi = -\beta e^{-i \phi}/2$}, \unexpanded{$\hat{K}_{\pm} = \hat{K}_x \pm i \hat{K}_y$}, 
%%%and \unexpanded{$| 0 \rangle$} is the vacuum.
%%%The state \unexpanded{$| \beta,\phi \rangle$} has coordinates 
%%%\unexpanded{$(K_x, K_y, K_z) \equiv (\langle \hat{K}_x \rangle, \langle \hat{K}_y \rangle, \langle \hat{K}_z \rangle)=
%%% \tfrac{1}{2}(\sinh \beta \cos \phi, \sinh \beta \sin \phi, \cosh \beta)$} 
%%%defining a hyperbolic surface. }
%%% 
%(see \new{right} panel in Fig.~\ref{fig1}).  
The interferometer protocol starts with vacuum in the $m_f=\pm1$ modes [Fig.~\ref{fig1}({I})]. 
The first~SMD $e^{-i \hat{H}_{\rm SMD} t/\hbar}$ [$(\chi t)_1 = \chi t$, $\phi_1=0$] generates a Lorentz boost 
\cite{LesliePRA2009, KlemptPRL2010}
that amplifies the population in the $m_f = \pm 1$ modes
\be \label{NumberSidesParam}
\mathcal{N}(t) = \frac{8\bar{n}^2}{4\bar{n}-1} \sinh^2 \Big( \frac{\sqrt{4\bar{n}-1}}{2} \chi t \Big),
\ee
where $\mathcal{N} \equiv \langle \hat{N}_{+1} + \hat{N}_{-1} \rangle_{\rm SMD}$ [Fig.~\ref{fig1}({II})]. 
The mean-field description is thus valid when 
\cite{Note2}
%%%\footnote{The quantum fidelity between \unexpanded{$|\psi_{1}\rangle= e^{-i 2 \chi t \bar{n} \hat{K}_x} |0, 0\rangle$} and
%%%\unexpanded{$|\psi_{2}\rangle = e^{-i \chi t ( \hat{a}_0^{\dag} \hat{a}_0^{\dag} \hat{a}_{+1} \hat{a}_{-1} + 
%%%  {\rm h.c.} )} | 0, \alpha, 0\rangle$},
%%%is \unexpanded{$\mathcal{F} = | \langle \psi_2 | \psi_1 \rangle | = | 1- \sum_{k=1}^{+\infty} (\chi\,t)^k\,f_k(\chi\,t\,\sqrt{\bar{n}}) |$}
%%%where \unexpanded{$f_k$} is a polynomial function with complex coefficients.
%%%We recover \unexpanded{$\mathcal{F} \approx 1 $} when 
%%%\unexpanded{$\chi t \to 0$}, \unexpanded{$\bar{n}\to \infty$} such that \unexpanded{$\chi t \sqrt{\bar{n}}$} remains finite. 
%%%Furthermore, according to Eq.~(\ref{NumberSidesParam}), requiring low depletion, 
%%%\unexpanded{$\mathcal{N}(t) \sim (\chi t \bar{n})^2 \ll \bar{n}$}, gives  
%%%the condition (\ref{Param_condition}).}
\be \label{Param_condition}
\chi t \to 0, \, \bar{n} \to +\infty, \qquad \text{such that} \,\,\, 0< \chi t \sqrt{\bar{n}} \ll 1.
\ee
The~SMD generates a thermal distribution of perfectly correlated atom pairs in the $\pm 1$ modes \cite{PuPRL2000,DuanPRL2000}: 
the two-mode squeezed-vacuum state \cite{ScullyBOOK}, with variance 
$(\Delta\hat N_{\pm1})_{\rm SMD}^2 = \frac{\mathcal{N}}{2}(\frac{\mathcal{N}}{2}+1)$.
The transformation $e^{-i\hat{H}_{\rm PS}t_{\rm PS}/\hbar}$ rotates the state around the $z$ axis of an angle $\theta$ [Fig.~\ref{fig1}({III})].
The final operation is a second~SMD. 
This can be implemented either as an inverse Lorentz boost $e^{i \hat{H}_{\rm SMD} t/\hbar}$ [i.e.~$(\chi t)_2 = -(\chi t)_1$, $\phi_2=0$, as in Fig.~\ref{fig1}({IV})], or
by applying a $\pi/2$ phase shift to the $m_f=0$ mode followed by the transformation $e^{-i \hat{H}_{\rm SMD} t/\hbar}$ [i.e.~$(\chi t)_2 = (\chi t)_1$, $\phi_2=\pi/2$].
The latter is easier to be realized experimentally~\cite{Note5}.
In both cases, the conditional probabilities are 
\be \label{Prob}
P(N_{\pm 1} \vert \theta) = \frac{2 [\mathcal{N}(\mathcal{N}+2)(1-\cos\theta)]^{N_{\pm1} }}{[\mathcal{N}(\mathcal{N}+2)(1-\cos\theta)+2 ]^{N_{\pm1}+1}}.
\ee
A direct calculation of Eq.~(\ref{Fisher}) yields
\be \label{FisherMeanField}
F(\theta) =\frac{\mathcal{N}(\mathcal{N}+2)}{ \mathcal{N}(\mathcal{N}+2) \sin^2 \frac{\theta}{2}+1}  \cos^2 \tfrac{\theta}{2},
\ee
where $\mathcal{N}$ is given by Eq.~(\ref{NumberSidesParam}).
The FI reaches its maximum at $\theta=0$. 
In this case, if $(\chi t)_2 = -(\chi t)_1$ the two~SMDs exactly compensate and the output $m_f=\pm 1$ modes are empty.
Note also that $\langle \hat{N}_{\pm 1} \rangle_{\rm out} = \mathcal{N}\big(\mathcal{N}+2\big)\sin^2\tfrac{\theta}{2}$ and 
$(\Delta \hat{N}_{\pm 1})^2_{\rm out} = (\Delta\hat{N})^2_{\rm SMD} \sin^2\tfrac{\theta}{2} [ (\Delta\hat{N})^2_{\rm SMD} \sin^2\tfrac{\theta}{2}+1 ]$:
error propagation 
%$\Delta \theta_{\rm ep} = \frac{(\Delta \hat{N}_{\pm 1})_{\rm out}}{ \vert\ud \langle \hat{N}_{\pm 1} \rangle_{\rm out}/\ud\theta \vert}$
saturates the Cram\'er-Rao lower bound, $\Delta \theta_{\rm ep} = \Delta \theta_{\rm CR}$.
At $\theta=0$, we obtain $\Delta \theta_{\rm CR}= 1/\sqrt{m\mathcal{N} (\mathcal{N}+2)}$, which 
is below the shot noise, $\Delta \theta_{\rm CR} < 1/\sqrt{m\mathcal{N}}$, calculated 
considering only the average population in $m_f=\pm 1$ after the first~SMD~\cite{YurkePRA1986, HudelistNATCOMM2014}.
We notice here that the shot noise should be calculated with respect to the total resources, i.e. the 
total average number of particles $\bar{n}$ in the input state.
However, such an analysis is impossible within the SU(1,1) framework.

%%%%%%%%%%%%%%%%%%%%%%%%%%%%%%%%%%%%%%%%%%%%%%%%%%%%%%%%%%%%%%%
% figure 2
%%%%%%%%%%%%%%%%%%%%%%%%%%%%%%%%%%%%%%%%%%%%%%%%%%%%%%%%%%%%%%%
\begin{figure}[!t] 
\includegraphics[width=0.485\textwidth]{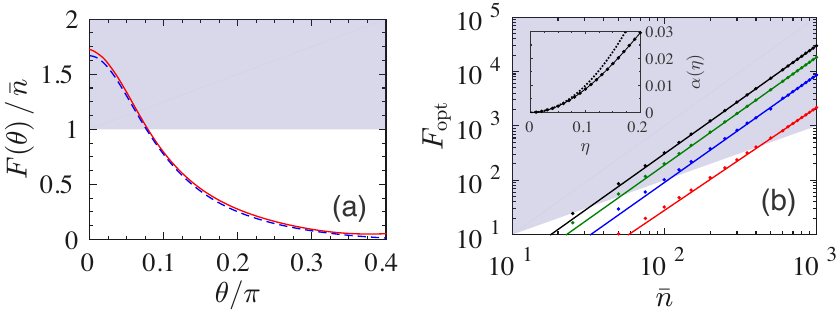}
\caption{%(color online) 
(a) The FI as a function of $\theta$ (solid red line) and error propagation $1/(\Delta\theta)^2_{\rm ep}$ (dashed blue line). % and \new{the quantum bound $F_Q$} (horizontal dashed green line). 
The shaded area is $F(\theta)> \bar{n}$. 
Here $\bar{n}=50$ and $\eta=0.2$.
(b) $F_{\rm opt}$ as a function of $\bar{n}$. 
Dots are numerical results and solid lines are quadratic fits to the data: $F_{\rm opt} = \alpha(\eta) \bar{n}^2$, for $\bar{n}\gg1$.
Here $\eta=0.05$ (red), $0.1$ (blue), $0.15$ (green) and $0.2$ (black).
The shaded area is $F_{\rm opt}> \bar{n}$.
The inset shows $\alpha(\eta)$ as a function of $\eta$ (dots). 
The solid line is a quadratic fit $\alpha(\eta) = \eta^2-1.31 \eta^3$; 
the dotted line is $\alpha(\eta) = \eta^2$.}
\label{fig2}
\end{figure}
%%%%%%%%%%%%%%%%%%%%%%%%%%%%%%%%%%%%%%%%%%%%%%%%%%%%%%%%%%%%%%%
%%%%%%%%%%%%%%%%%%%%%%%%%%%%%%%%%%%%%%%%%%%%%%%%%%%%%%%%%%%%%%%
%%%%%%%%%%%%%%%%%%%%%%%%%%%%%%%%%%%%%%%%%%%%%%%%%%%%%%%%%%%%%%%

{\it Full quantum approach.}~We have thus performed a full three-mode quantum analysis, 
investigating the regime of parameters beyond Eq.~(\ref{Param_condition}).
Thanks to the symmetry of the Hamiltonian~(\ref{Hamiltonian}), 
we can restrict ourselves to the Hilbert subspace spanned by Fock states 
$\{ \vert N_{-1}, N_0, N_{+1} \rangle \equiv \vert k, M-2k, k \rangle \}$, with $0\leq k\leq \left\lfloor\frac{M}{2}\right\rfloor$ \cite{MiasPRA2008, GoldstonePRA1999}. 
We take 
$\hat{\rho} = \sum_{M=0}^{+\infty} \frac{\bar{n}^M e^{ -\bar{n} }}{M!  } \vert 0,M,0\rangle \langle 0, M, 0 \vert$ as the (input) probe state.

We numerically calculate $F(\theta)$ for different values of the parameters $\bar{n}$, $\eta$ and $\theta$, where 
$\eta \equiv \langle \hat{N}_{+1} + \hat{N}_{-1} \rangle_{\rm SMD}/\bar{n} $ is the fraction of particles transferred from the $m_f=0$ mode to the $m_f=\pm 1$ modes after the first~SMD.
We mainly focus on the case $(\chi t)_2  = -(\chi t)_1$ which, as shown below, is optimal.
Overall, the FI as a function of $\theta$ shows a behavior qualitatively similar to Eq.~(\ref{FisherMeanField}), with
a maximum at $\theta=0$, see Fig.~\ref{fig2}(a). 
A first important result is that, for proper values of $\eta$, 
the FI can be larger than $\bar{n}$ or, equivalently, $\Delta \theta_{\rm CR} < 1/\sqrt{m\bar{n}}$. 
In other words, it is possible to attain SSN uncertainties with respect to the average input number of particles.
\begin{figure}[!t] 
\includegraphics[width=0.48\textwidth]{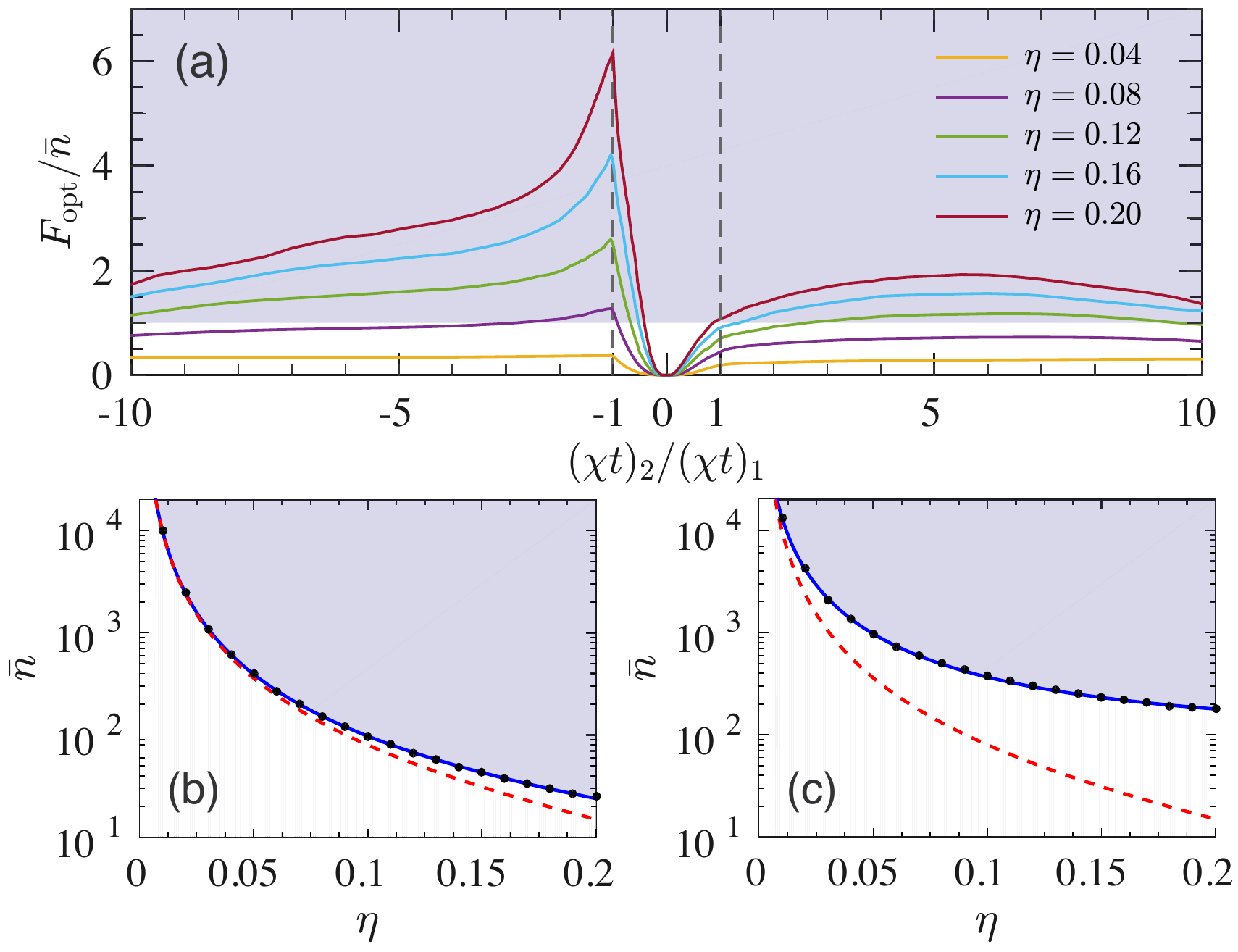}
\caption{%(color online) 
(a) $F_{\rm opt}$ as a function of the ratio $(\chi t)_2/(\chi t)_1$.
Solid lines refer to different values of $\eta$. Here $\bar{n}=200$.
Panels (b) and (c) show the phase-sensitivity portrait in the $(\eta,\bar{n})$-parameter space
for $(\chi t)_1 = -(\chi t)_2$ and $(\chi t)_1 = (\chi t)_2$, respectively.
SSN phase uncertainties are obtained for $\bar{n}$ 
larger than a critical value $\bar{n}_{\rm cr}(\eta)$ (dots, the solid line being a guide to the eye).
The dashed red line in both panels is $\bar{n}_{\rm cr}(\eta) = (1-2\eta)/\eta^2$, obtained from a mean-field calculation, which
agrees with the numerics in the limit (\ref{Param_condition}). In all panels, the shaded area indicates SSN.}
\label{fig3}
\end{figure}
%%%%%%%%%%%%%%%%%%%%%%%%%%%%%%%%%%%%%%%%%%%%%%%%%%%%%%%%%%%%%%%
%%%%%%%%%%%%%%%%%%%%%%%%%%%%%%%%%%%%%%%%%%%%%%%%%%%%%%%%%%%%%%%
%%%%%%%%%%%%%%%%%%%%%%%%%%%%%%%%%%%%%%%%%%%%%%%%%%%%%%%%%%%%%%%

A scaling analysis of the FI as a function of $\bar{n}$ at the optimal point $\theta=0$ [we indicate $F_{\rm opt} \equiv \max_\theta F(\theta)$]
shows that $F_{\rm opt} \approx \alpha(\eta) \bar{n}^2$ asymptotically in $\bar{n}$ (in our simulations $\bar{n} \lesssim 1000$), see Fig.~\ref{fig2}(b).
A fit gives $\alpha(\eta) \approx \eta^2 (1 -1.3\eta)$ in the case $(\chi t)_2/(\chi t)_1 = -1$ [see the inset of Fig.~\ref{fig2}(b)].
We thus conclude that $\Delta \theta_{\rm CR} \sim 1/\bar{n}$ with a prefactor depending on $\eta$.

Figure~\ref{fig3} is the main result of this Letter. 
In panel (a) we show $F_{\rm opt}$ as a function of the ratio $(\chi t)_2/(\chi t)_1$, for different values of $\eta$. 
For relatively large $\eta$, outside the mean-field regime, the curves are asymmetric around zero.
The optimal interferometer configuration is reached for $(\chi t)_2=-(\chi t)_1$, 
but SSN can be also obtained for positive values of $(\chi t)_1/(\chi t)_2$:
inverting the sign of $\chi$ in the second~SMD transformation, which might be experimentally difficult, is not necessary to reach SSN sensitivities.
Figures~\ref{fig3}(b) and \ref{fig3}(c) show the regime of parameters where SSN can be achieved, 
for $(\chi t)_2=- (\chi t)_1$ and $(\chi t)_2= (\chi t)_1$, respectively.
For fixed $\eta$, a critical value $\bar{n}_{\rm cr}(\eta)$ exists 
such that $\Delta \theta_{\rm CR} \leq 1/\sqrt{m \bar{n}}$, for $\bar{n} > \bar{n}_{\rm cr}(\eta)$.
%In the mean-field limit 
%$\bar{n}_{\rm cr}(\eta) =  (1-2\eta)/\eta^2$,
%which agrees with the numerics in the limit (\ref{Param_condition}). 
Deviations from the mean-field prediction, $\bar{n}_{\rm cr}(\eta) =  (1-2\eta)/\eta^2$, can be appreciated for small $\bar{n}$, especially for $(\chi t)_2/(\chi t)_1>0$, 
and are relevant in current BEC experiments \cite{GrossNATURE2011, HamleyNATPHYS2012,GervingNATCOMM2013, LuckeSCIENCE2011}.

\begin{figure}[!t] 
\includegraphics[width=0.485\textwidth]{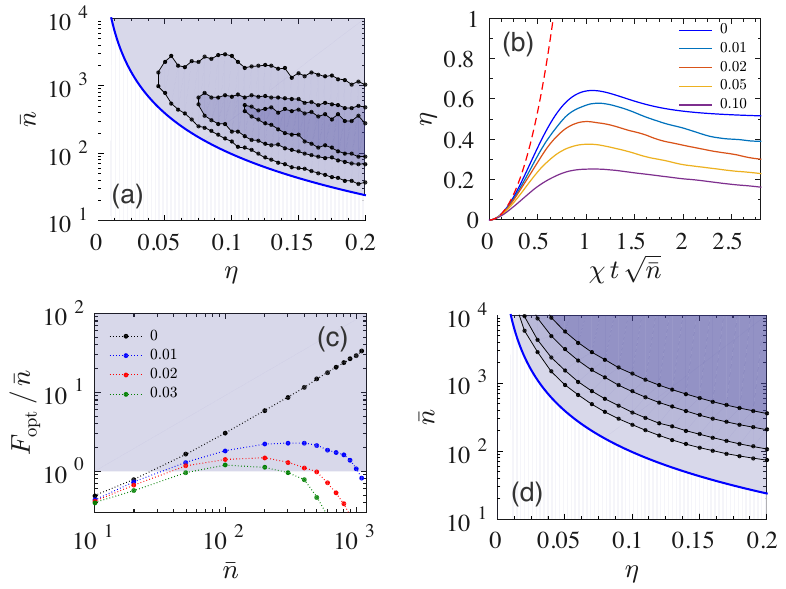} 
\caption{%(color online) 
(a) SSN region in $(\eta,\bar{n})$-parameter space including two-body losses in the $m_f=0$ mode, 
with loss parameter $\gamma/\chi=0.01$, $0.02$ and $0.03$ (from outer to inner regions).
The thick blue line, for $\gamma=0$, is the same as the solid line in Fig.~\ref{fig3}.
(b) $\eta$ as a function of time, for different values of $\gamma/\chi$ (solid lines).
Here $\bar{n}=200$ and the dashed line is Eq.~(\ref{Param_condition}).
(c) $F_{\rm opt}$ as a function of $\bar{n}$ for $\eta=0.2$ and different values of $\gamma/\chi$.
Dotted lines are guides to the eye.
(d) SSN region  in $(\eta,\bar{n})$-parameter space obtained for detection noise $\sigma=1$, $2$, $5$, $10$ (from outer to inner regions). 
The thick blue line ($\sigma=0$) corresponds to the solid line in Fig.~\ref{fig3}.
In panels (a) and (d) the dots are numerical results and the solid lines are guides to the eye.
}
\label{fig4}
\end{figure}

{\it Particle loss and finite detection efficiency.}~According to Eq.~(\ref{NumberSidesParam}), the~SMD is unaffected by decoherence processes that happen 
on time scales much longer than $ \sim 1/(\chi \sqrt{\bar{n}})$. 
%For sufficiently large $\bar{n}$ and fast phase encoding, the nonlinear interferometer thus appears to be robust to one-body losses
%(relevant for the spin-mixing dynamics in the $f=1$ manifold  \cite{Peise2015}).
For sufficiently large $\bar{n}$ and fast phase encoding, the nonlinear interferometer thus appears to be robust 
to one-body losses (relevant for the spin-mixing dynamics in the $f=1$ manifold~\cite{Peise2015}). 
%\new{These are density-independent and due, for instance, to inelastic collisions of the ultracold trapped atoms with the background thermal cloud
%or scattering by off-resonant light in a dipole trap.}
In fact, this dissipation source 
-- due to inelastic collisions of the ultracold trapped atoms with the background thermal cloud, or by off-resonant light scattering in a dipole trap -- 
has a density-independent rate.
Conversely, recombination losses -- whose rate depends on $\bar{n}$ -- 
may strongly affect the interferometer sensitivity.
We have thus simulated two-body losses in the $m_f=0$ mode
(relevant for the spin-mixing dynamics in the $f=2$ manifold  \cite{GrossNATURE2011, LuckeSCIENCE2011})
using a Monte Carlo wave-function approach \cite{MolmerJOSAB1993,*LiPRL2008}.
Let $\gamma$ indicate the depletion rate during the~SMD operation [i.e. $\langle \hat{N}_0(t) \rangle = \bar{n} / (1 + 2\gamma t \bar{n})$ for $\chi=0$].
Figure~\ref{fig4}(a) shows the regime of parameters $(\eta,\bar{n})$ where SSN sensitivities can be found. 
The SSN region shrinks when increasing $\gamma/\chi$ and, in particular, no SSN is found for $\gamma/\chi \gtrsim 0.04$. 
The branch structure of the SSN regions is explained by the characteristic effects induced by particle losses shown in 
Figs.~\ref{fig4}(b) and \ref{fig4}(c). 
In Fig.~\ref{fig4}(b) we plot $\eta$ as a function of time, for different values of $\gamma/\chi$.
Losses decrease the transfer rate and place an upper bound to the achievable $\eta$.
In Fig.~\ref{fig4}(c) we show the FI as a function of $\bar{n}$. 
For $\bar{n} \ll (\chi/2\gamma)^2$, the effect of losses can be neglected and we 
recover the scaling $F_{\rm opt} \propto (\eta \bar{n})^2$ of the noiseless case. 
For $\bar{n} \gtrsim (\chi/2\gamma)^2$ losses dominate and the sensitivity quickly degrades. 
For instance, in typical experiments with $^{87}$Rb in the $f=2$ manifold, the coupling strength is $\chi \approx 0.5$ Hz
and we estimate a ratio $\gamma / \chi \approx 10^{-3} - 10^{-2}$, well within our explored range.

%%%%%%%%%%%%%%%%%%%%%%%%%%%%%%%%%%%%%%%%%%%%%%%%%%%%%%%%%%%%%%%
% figure 4
%%%%%%%%%%%%%%%%%%%%%%%%%%%%%%%%%%%%%%%%%%%%%%%%%%%%%%%%%%%%%%%
%%%%%%%%%%%%%%%%%%%%%%%%%%%%%%%%%%%%%%%%%%%%%%%%%%%%%%%%%%%%%%%
%%%%%%%%%%%%%%%%%%%%%%%%%%%%%%%%%%%%%%%%%%%%%%%%%%%%%%%%%%%%%%%
%%%%%%%%%%%%%%%%%%%%%%%%%%%%%%%%%%%%%%%%%%%%%%%%%%%%%%%%%%%%%%%

To model finite detection efficiency we consider a Gaussian convolution of the ideal output probabilities
\cite{note4,PezzePRL2013}. 
Results for different values of the detection noise $\sigma$ are shown in Fig.~\ref{fig4}(d).
In typical experiments $\sigma \approx 10$, while a high detection sensitivity $\sigma \approx 1$ has been discussed in Ref.~\cite{MusselPRL2014}.
In the regime (\ref{Param_condition}) we can evaluate the FI from a convolution of probabilities~(\ref{Prob}). 
This allows for semianalytical calculations giving, to the leading order in $1/\eta$ and for $\sigma \gtrsim 1$,
$\bar{n}_{\rm cr}(\eta) \approx 2 \sigma / \eta^2$, 
which agrees with numerical calculations for $\bar{n} \to +\infty$ and $\eta \to 0$.
It predicts that $n_{\rm cr}(\eta)$ shifts toward larger values when increasing $\sigma$, 
an expected behavior~\cite{MarinoPRA2012} that
qualitatively holds also outside the mean-field regime. 

{\it Conclusions.}~We have studied a nonlinear three-mode interferometer with spinor BECs. 
The nonlinear spin-mixing dynamics not only splits the initial cloud but, differently from a linear beam splitter, 
it also creates, at the same time, quantum correlations among particles, necessary to overcome the shot-noise limit. Therefore,
differently from linear interferometers, the nonlinear scheme discussed in this Letter 
can reach SSN phase uncertainties with classically correlated probe states. 
%Quantum correlations necessary to overcome the shot-noise limit are created by the nonlinear spin mixing dynamics.
Accurate predictions of the phase sensitivity require a full three-mode quantum analysis, beyond the SU(1,1) (mean-field) approach. 
We have performed such an analysis and showed that it is possible to overcome the shot-noise limit with respect to the total average number of atoms in input.
We also provide the regime of parameters where sub-shot-noise uncertainties can be achieved, including 
losses and finite detection efficiencies.
Our results pave the way to atomic ultrasensitive spin-mixing interferometry~\cite{Note5}.
%%%\footnote{Preliminary  experimental results can be found in \new{the} Master's theses of 
%%J.~Schulz, and D.~Linnemann, \new{both} available at 
%%http://www.kip.uni-heidelberg.de/matterwaveoptics/publications/theses/.}.
%%%

\begin{acknowledgements}
{\it Acknowledgements.} We thank C. Klempt, B. L\"ucke, W. M\"ussel, M.K. Oberthaler and H. Strobel for discussions. 
This work is supported by EU-STREP Project QIBEC, No. FP7-ICT-2011-C.
LP acknowledges financial support by MIUR through FIRB Project No. RBFR08H058.
\end{acknowledgements}

%%\bibliography{biblio}

\begin{thebibliography}{56}
\bibitem [{Note4()}]{note4}%
  \BibitemOpen
  \bibinfo {note} {The conditional probabilities in presence of detection noise
  are $P_{\protect \rm dn}(n | \theta ) = \DOTSB \sum@ \slimits@
  _{n'=0}^{+\infty } P(n' | \theta ) \protect \mathcal {C}(n')
  e^{-(n-n')^2/2\sigma ^2}$, where $n=N_{\pm 1}$, $\protect \mathcal {C}(n')$
  provides the normalization, $\sigma $ is number-independent, and $P(n |
  \theta )$ are ideal (noiseless) probabilities. The FI is evaluated by
  replacing $P(n | \theta )$ with $P_{\protect \rm dn}(n | \theta )$ in
  Eq.~(\ref {Fisher}).}\BibitemShut {Stop}%
\end{thebibliography}
%%\bibliographystyle{apsrev4-1}

%merlin.mbs apsrev4-1.bst 2010-07-25 4.21a (PWD, AO, DPC) hacked
%Control: key (0)
%Control: author (72) initials jnrlst
%Control: editor formatted (1) identically to author
%Control: production of article title (-1) disabled
%Control: page (0) single
%Control: year (1) truncated
%Control: production of eprint (0) enabled
%

\end{document}